\documentclass[10pt,twocolumn]{article}
\usepackage[a4paper,margin=0.75in,columnsep=0.25in]{geometry}
\usepackage[T1]{fontenc}
\usepackage{newtxtext,newtxmath}
\usepackage{microtype}
\usepackage{graphicx}
\usepackage{booktabs}
\usepackage{array}
\usepackage[table]{xcolor}
\usepackage{amsmath}
\usepackage{tikz}
\usetikzlibrary{arrows.meta,positioning,fit,calc}
\usepackage[hidelinks]{hyperref}
\usepackage[capitalise,noabbrev]{cleveref}
\usepackage{caption}
\usepackage{enumitem}
\setlist{nosep,leftmargin=*}
\usepackage{natbib}
\setcitestyle{numbers,square}

\definecolor{ink}{HTML}{0B0B0B}
\definecolor{ink2}{HTML}{52514E}
\definecolor{sblue}{HTML}{2A78D6}
\definecolor{rowhl}{HTML}{EAF2FC}
\newcommand{\bthree}{B$^3$-F1}

\title{\textbf{Line-Level Singer Diarization in Multi-Member Pop Songs:}\\[2pt]
\textbf{A Benchmark, Lead-Vocal Isolation, and Song-Level Decoding}}
\author{Allen Ng\\[2pt]
\normalsize nagiflow LLC, Tokyo, Japan\\
\normalsize\texttt{allen.ng@nagiflow.com}}
\date{}

\begin{document}
\maketitle

\begin{abstract}
Fans of multi-member pop groups annotate songs line by line with the member who sings each line, but speech
diarization does not transfer to this setting: a commercial diarizer finds one or two speakers in songs by groups of
four to eight. We study \emph{line-level singer diarization}: given a song and its lyric-line timings, group the
lines by singer with no prior knowledge of the group. We release \textbf{ColorCoded-SD}, a benchmark of 43 K-pop
songs (1,740 lines, 3--8 singers per song) built from crowd annotations verified by independent annotators, together
with alignment and scoring tools, and we derive line labels for the IdolSongsJp corpus from its per-singer solo
renditions for evaluation on unseen singers. Combining lead-vocal isolation with a karaoke separator,
singing-adapted speaker embeddings, and two training-free song-level decoders (a sticky Viterbi pass over line order
and unsupervised linking of a group's songs) improves B-cubed F1 over the standard pipeline (vocal separation, speaker
embeddings, per-song clustering) by +0.058 on ColorCoded-SD ($p=0.001$) and by +0.066 on IdolSongsJp ($p=0.003$).
Isolating the lead is the largest single factor wherever the separator leaves a lead voice (+0.04 to +0.06 on the
same lines), and with the singers' dry studio takes two different encoders become indistinguishable; the remaining
errors concentrate on short lines that carry too little audio for a voiceprint. We also report negative results
(more or cleaner training data, weight editing, a song-context model, and target-singer extraction) that bound
where further gains can come from.
\end{abstract}

\section{Introduction}
\label{sec:intro}

K-pop and J-pop idol groups have four to more than ten members who trade lines within a song. A popular fan format,
the \emph{line-distribution} or \emph{color-coded lyrics} video, assigns each lyric line a member color; fans make
these by hand, and the annotation tool we draw our data from holds 19,042 such projects for 6,241 artists. Deciding
who sings each line is the slowest step and a natural target for speaker diarization, but diarization built for
speech does not transfer: on full mixes, a current commercial service~\citep{elevenlabs2025scribe} finds on average
1.35 speakers in songs with 5.3 singers, and a single speaker in 33 of 43 songs (\cref{sec:commercial}).

The task differs from speech diarization in three ways. (i) \emph{Line timings} are known from the lyrics or forced
alignment, so the problem is grouping, not segmentation. (ii) Members of a group are chosen to blend: they share a
vocal style, a production and often a pitch range, and their lines are layered with harmonies and ad-libs from other
members. (iii) The number of singers is usually known from the member list; in 95.7\% of 6,915 annotated projects,
every member created by the annotator sings in the song. We therefore evaluate with a known singer count and report
count estimation separately.

We find that the grouping error decomposes into three parts of different size. Where a karaoke separator leaves the
lead voice, \emph{isolating it} is the largest single factor (+0.04 to +0.06 \bthree{} on the same lines), larger than
any change of speaker encoder; on IdolSongsJp, two encoders that differ by 0.11 \bthree{} on the full mix become
identical on the singers' dry takes. \emph{Song-level decoding}, which exploits that singer turns persist and that a
group's members recur across its songs, adds +0.028 on top. The \emph{remaining} errors sit mostly on short lines
that carry too little audio for any voiceprint.

\paragraph{Contributions.}
\begin{itemize}
  \item \textbf{ColorCoded-SD}, a public line-level singer-diarization benchmark of 43 K-pop songs with confirmed and
  single-annotator lines, a measured human agreement of 89.1\%, and alignment and scoring tools; and a procedure that
  derives line labels for IdolSongsJp from its solo versions (\cref{sec:data}).
  \item \textbf{A decomposition} of the error into input, encoder and decoding, with paired song-bootstrap tests for
  every comparison and one scorer for every number (\cref{sec:results}).
  \item \textbf{Two training-free song-level decoders} that together add +0.028 \bthree{} ($p<0.001$) for the same
  encoder and input (\cref{sec:decoding}).
  \item \textbf{A singing-adaptation recipe} for speaker encoders (in-song contrastive negatives and an anchor to the
  speech model) that lets a 15M-parameter ResNet match ECAPA2 at 1/8 of its CPU cost, and \textbf{negative results}
  that bound the remaining headroom (\cref{sec:negative}).
\end{itemize}

\section{Related Work}
\label{sec:related}

\paragraph{Speaker diarization.} Modern diarization combines speaker embeddings with clustering, often refined by an
HMM over time (VBx~\citep{landini2022vbx}) or replaced by end-to-end models~\citep{plaquet2023powerset};
see~\citep{park2022review} for a review. TS-VAD~\citep{medennikov2020tsvad} uses embeddings from a first clustering
pass as targets for a second pass. Our sticky decoder is a lightweight VBx-style transition prior over \emph{lines}
rather than frames.

\paragraph{Singer diarization and idol-group corpora.} FruitsMusic~\citep{suda2024fruitsmusic} introduced singer
diarization for Japanese idol-group music videos and showed that ECAPA-TDNN~\citep{desplanques2020ecapa} embeddings
degrade on mixtures. IdolSongsJp~\citep{suda2025idolsongsjp} provides commissioned idol-style songs with per-singer
recordings. FruitsMusic clusters a group's lines across several songs and scores frames; we cluster within each song
and score lines, so numbers are not directly comparable (\cref{sec:limits}).

\paragraph{Source separation.} HT-Demucs~\citep{rouard2023htdemucs}, BS-RoFormer~\citep{lu2024bsroformer} and
Mel-RoFormer~\citep{wang2023melroformer} separate vocals from accompaniment; community ``karaoke'' models further
split lead from backing vocals~\citep{becruily2025karaoke}. Singer-informed separation~\citep{xu2026singerinformed}
conditions a separator on a singer embedding from a clean enrollment clip for two-singer duets. We use separators as
fixed front ends and measure their effect on grouping.

\paragraph{Speaker embeddings for singing.} We build on ECAPA2~\citep{thienpondt2023ecapa2} and the SpeechBrain
ResNet~\citep{ravanelli2021speechbrain}. Self-supervised speech and music models (WavLM~\citep{chen2022wavlm},
MERT~\citep{li2024mert}) and a VoxCeleb ResNet-293~\citep{wang2023wespeaker} were clearly worse in our preliminary
comparison (leave-one-out centroid accuracy $\le$72.5\% against 75.9--78.7\% for the two speech encoders).

\section{Task and Data}
\label{sec:data}

\paragraph{Task.} A song has lines $\ell=1,\dots,N$ in time order with known start and end times. Each
\emph{single-singer} line has one singer $y_\ell\in\{1,\dots,k\}$. A system outputs a partition
$z_\ell\in\{1,\dots,k\}$ of all lines; singer identities are unknown, and no other song of the group is used unless
stated (\cref{sec:decoding}). Lines sung by several members (16.4\% of annotated lines) are not part of the task.

\paragraph{ColorCoded annotations.} Annotations come from user projects of a line-coloring web application (19,042
projects, 6,241 artists). For evaluation we use five groups, BLACKPINK (4 members), ILLIT (5), KATSEYE (6),
BABYMONSTER (7) and Stray Kids (8), and keep a song only if projects by at least two different users agree on who
sings when (75\% agreement on a 0.25\,s grid). This removes fan covers (``how would X sing this'') and careless
projects, and leaves 50 verified songs.

\paragraph{ColorCoded-SD.} The released benchmark contains the 43 verified songs whose annotated audio matches the
official recording (7 fan edits and excerpts are withheld). For each song we keep the lines of the best-agreeing
project and drop any line that another user's time-aligned project assigns to a different singer (30 lines). Lines
that another project confirms are marked \emph{confirmed} (1,470); lines that no other project covers are kept as
\emph{single-annotator} lines (270). Timings are snapped to a 0.25\,s grid and singers are pseudonymous. Because the
songs and timings refer to the annotators' audio copies, the release includes official recording identifiers, 10\,Hz
chroma fingerprints, and a DTW script that maps the timings onto a user's own copy (in a shift-and-trim test it
recovers line times to within 0.03\,s).

\paragraph{Human agreement.} 319 songs in the full collection have two independent projects with at least six
aligned single-singer lines (12,914 lines). After one-to-one mapping of member names the annotators agree on
\textbf{89.1\%} of lines (median song 100\%). We use this as a reference level for one-to-one line accuracy.

\paragraph{IdolSongsJp labels.} IdolSongsJp~\citep{suda2025idolsongsjp} contains 15 commissioned songs by 18 session
singers, none of whom appear in our data. Each singer recorded the whole song and the master uses one take per line,
so the dry tracks alone cannot label lines. The corpus also contains, for every singer $X$, a \emph{solo} version:
the same mix with all vocals replaced by $X$. Wherever $X$ sings the lead in the master, master $\approx$ solo$_X$. On
100\,ms frames we compute $R_X=\lVert\text{master}-\text{solo}_X\rVert^2/\lVert\text{master}\rVert^2$ and label a frame
single-singer if $\min_X R_X<0.35$ and the best residual is below $0.35\times$ the second best; labelled runs of at
least 1\,s form lines. This yields 309 lines in 14 songs (one song is fully layered), covering 45\% of vocal time.
IdolSongsJp is used \emph{only} for evaluation; nothing was trained or tuned on it.

\paragraph{Training corpus.} Encoder adaptation uses projects of other artists (all evaluation artists, including
solo and mash-up projects of their members, are excluded): 2,277 songs, 1,765 singers, 364 artists, about 125\,h of
singing, labelled by single unverified projects.

\begin{table}[t]
\centering\small
\caption{Data. Every result in this paper is measured on ColorCoded-SD or IdolSongsJp.}
\label{tab:data}
\setlength{\tabcolsep}{3pt}
\begin{tabular}{@{}lrrrl@{}}
\toprule
 & Songs & Singers & Lines & Use \\
\midrule
\textbf{ColorCoded-SD} & 43 & 30 & 1,740$^a$ & benchmark \\
IdolSongsJp & 14 & 18 & 309 & test only \\
Training corpus & 2,277 & 1,765 & --- & adaptation \\
Human agreement & 319 & --- & 12,914 & reference \\
\bottomrule
\end{tabular}\\[2pt]
{\footnotesize $^a$1,470 confirmed by a second annotator; 3--8 singers per song.}
\end{table}

\section{Method}
\label{sec:method}

\begin{figure*}[t]
\centering
\begin{tikzpicture}[
  font=\footnotesize, node distance=4mm,
  box/.style={draw=ink2, line width=0.5pt, rounded corners=2pt, fill=white, align=center, minimum height=10mm, inner sep=3pt},
  hl/.style={box, draw=sblue, line width=0.9pt, fill=rowhl},
  arr/.style={-{Stealth[length=2mm]}, line width=0.6pt, color=ink2}]
\node[box] (song) {Song audio\\+ line timings};
\node[hl, right=of song] (sep) {\textbf{Lead-vocal}\\\textbf{separation}\\(karaoke)};
\node[box, right=of sep] (win) {3\,s windows\\1.5\,s hop,\\silence gate};
\node[box, right=of win] (enc) {Speaker encoder\\(ECAPA2-ft /\\ResNet-ft)};
\node[box, right=of enc] (norm) {Song centering,\\line mean,\\$\ell_2$ norm};
\node[box, right=of norm] (ahc) {AHC\\($k$ clusters)};
\node[hl, right=of ahc] (dec) {\textbf{Song-level}\\\textbf{decoding}\\sticky + discography};
\node[box, right=of dec] (out) {Line\\$\to$ singer\\cluster};
\foreach \a/\b in {song/sep, sep/win, win/enc, enc/norm, norm/ahc, ahc/dec, dec/out} \draw[arr] (\a) -- (\b);
\node[below=1.5mm of sep, text=ink2, font=\scriptsize] {\cref{sec:input}};
\node[below=1.5mm of enc, text=ink2, font=\scriptsize] {\cref{sec:encoders}};
\node[below=1.5mm of dec, text=ink2, font=\scriptsize] {\cref{sec:decoding}};
\end{tikzpicture}
\caption{Pipeline. The highlighted stages carry the gain over the standard pipeline (vocal separation, speaker
encoder, per-song clustering).}
\label{fig:pipeline}
\end{figure*}
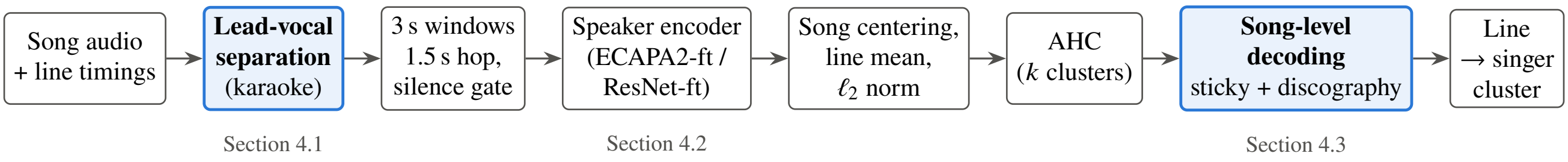

\Cref{fig:pipeline} shows the pipeline. Audio is separated (\cref{sec:input}) and resampled to 16\,kHz; each line is
trimmed by 0.1\,s at both ends and cut into 3\,s windows with 1.5\,s hop, and windows whose mean absolute amplitude is
below a quarter of the song's mean are dropped. An encoder maps each window to $x_w$. Within a song we subtract the
song's mean embedding and $\ell_2$-normalize, $\tilde x_w=\mathrm{norm}(x_w-\bar x_{\text{song}})$, and represent line
$\ell$ by $v_\ell=\mathrm{norm}\big(\tfrac{1}{|W_\ell|}\sum_{w\in W_\ell}\tilde x_w\big)$; centering removes the
production component shared by all singers of a song. Lines are clustered by average-linkage agglomerative clustering
(AHC) on cosine distance into $k$ clusters. A line shorter than 1.2\,s, or silent after separation, has no window and
hence no voiceprint (11\% of ColorCoded-SD lines); it takes the cluster of the nearest line in time that has one.

\subsection{Lead-vocal input}
\label{sec:input}
Besides the full mix and HT-Demucs~\citep{rouard2023htdemucs} vocals, we use a Mel-RoFormer karaoke
model~\citep{becruily2025karaoke} that keeps the lead and removes backing vocals and harmonies (overlap 2). A karaoke
model returns silence for lines it judges to be backing vocals; for those lines the voiceprint falls back to the
Demucs track, which is better than leaving them without one (0.699 vs.\ 0.661 \bthree{} for the full system).

\subsection{Singing-adapted speaker encoders}
\label{sec:encoders}
Speech-trained encoders confuse members of the same group. We adapt them on the training corpus with supervised
contrastive learning~\citep{khosla2020supcon} (temperature 0.1): positives are windows of the same singer in any song,
and negatives from a \emph{different singer of the same song} are weighted $\times2$, because that is exactly the
confusion grouping must resolve. An anchor term keeps the speech knowledge,
\begin{equation}
\mathcal{L}=\mathcal{L}_{\text{SupCon}}^{\times2\,\text{in-song}}+\lambda\,\big(1-\cos(z,z_{0})\big),\quad\lambda=0.2,
\end{equation}
with $z_0$ the frozen original embedding. Batches hold 3 singers $\times$ 3 windows from each of 3--6 songs plus one
window of each singer from another song; BatchNorm statistics stay frozen; augmentation is random gain and smooth
$\pm6$\,dB equalization. Architectures are unchanged: no layers or parameters are added and the inference cost is the
same. \textbf{ResNet-ft} adapts the SpeechBrain ResNet (15.4M parameters; layers 3--4 and the head trainable; AdamW,
lr $10^{-4}$, 1,000 steps) and \textbf{ECAPA2-ft} adapts ECAPA2 (27.1M; the 1D part trainable; lr $5\cdot10^{-5}$,
500 steps). Both train on a 16\,GB laptop GPU.

\subsection{Song-level decoding}
\label{sec:decoding}
\paragraph{Sticky decoding.} Consecutive single-singer lines keep the same singer 57.2\% of the time in 290,116
annotated line pairs, against $\approx1/k\approx20\%$ by chance. Starting from AHC we alternate centroid estimation
$c_j=\mathrm{norm}(\sum_{z_\ell=j}v_\ell)$ and a Viterbi pass that maximizes
\begin{equation}
\textstyle\sum_{\ell}\cos(v_\ell,c_{z_\ell})+\beta\sum_{\ell>1}\mathbb{1}[z_\ell=z_{\ell-1}]
\end{equation}
until the labels stop changing.

\paragraph{Discography linking.} Members of a group recur across its songs. Without labels we (1) cluster each song
into $k_s$ centroids $C_s$; (2) link each song's centroids one-to-one to $R$ group identities (the roster size) with
the Hungarian method~\citep{kuhn1955hungarian}, re-estimating identity prototypes $P_r$ as the normalized mean of their
linked centroids until convergence; and (3) re-decode each song against the blended references
$\mathrm{norm}(a\,P_{\pi_s(j)}+(1-a)\,C_{s,j})$ for five rounds, updating $C_s$ from the new labels and using the
sticky Viterbi pass as decoder. We use $a=0.75$ and $\beta=0.2$, selected leave-one-group-out on the ColorCoded songs.

\section{Evaluation Protocol}
\label{sec:protocol}
Every labelled line is predicted and scored with the scorer released with ColorCoded-SD. We report B-cubed
F1~\citep{bagga1998bcubed,amigo2009comparison} (per-line precision and recall, micro-averaged over lines: ``are lines
grouped correctly'') and one-to-one line accuracy after Hungarian mapping of clusters to singers (``Acc.'', what a
user sees after naming each cluster once). The singer count $k$ is given (\cref{sec:intro}). Intervals are 95\%
percentiles over 2,000 bootstrap resamples of \emph{songs}; comparisons resample the same songs for both systems and
report two-sided $p$-values. All components were developed and tuned on ColorCoded songs, so IdolSongsJp is the only
fully untouched test.

\section{Results}
\label{sec:results}

\begin{table*}[t]
\centering\small
\caption{Main results (true $k$, every line predicted). \bthree{} with 95\% song-bootstrap CI and one-to-one line
accuracy (Acc., \%). $\Delta$B$^3$: paired difference to the standard pipeline (ECAPA2 on Demucs vocals);
$^{\ast}$: $p<0.05$. Lead: karaoke lead vocals, Demucs voiceprint where the lead track is silent.}
\label{tab:main}
\setlength{\tabcolsep}{4pt}
\begin{tabular}{@{}llccc@{\hspace{6pt}}cc@{\hspace{6pt}}ccc@{}}
\toprule
 & & \multicolumn{5}{c}{ColorCoded-SD (43 songs)} & \multicolumn{3}{c}{IdolSongsJp (14 songs, unseen)} \\
\cmidrule(lr){3-7}\cmidrule(l){8-10}
 & & \multicolumn{3}{c}{all 1,740 lines} & \multicolumn{2}{c}{confirmed 1,470} & \multicolumn{3}{c}{all 309 lines} \\
\cmidrule(lr){3-5}\cmidrule(lr){6-7}\cmidrule(l){8-10}
System & Input & B$^3$-F1 [95\% CI] & Acc. & $\Delta$B$^3$ & B$^3$-F1 & $\Delta$B$^3$ & B$^3$-F1 [95\% CI] & Acc. & $\Delta$B$^3$ \\
\midrule
ECAPA2 [standard] & Demucs & 0.641 {\scriptsize[0.58,0.70]} & 66.3 & --- & 0.651 & --- & 0.800 {\scriptsize[0.72,0.88]} & 81.6 & --- \\
ECAPA2 & lead & 0.661 {\scriptsize[0.60,0.72]} & 68.6 & +0.020 & 0.672 & +0.020 & 0.847 {\scriptsize[0.76,0.92]} & 85.8 & +0.047 \\
ResNet-ft & lead & 0.665 {\scriptsize[0.60,0.73]} & 69.0 & +0.024 & 0.657 & +0.006 & 0.824 {\scriptsize[0.75,0.90]} & 82.8 & +0.024 \\
ECAPA2-ft & lead & 0.672 {\scriptsize[0.61,0.73]} & 69.3 & +0.031$^{\ast}$ & 0.673 & +0.022 & 0.859 {\scriptsize[0.78,0.93]} & 87.1 & +0.059$^{\ast}$ \\
\quad + sticky decoding & lead & 0.677 {\scriptsize[0.61,0.74]} & 69.7 & +0.036$^{\ast}$ & 0.680 & +0.028 & 0.875 {\scriptsize[0.79,0.94]} & 87.7 & +0.075$^{\ast}$ \\
\quad + discography linking & lead & 0.678 {\scriptsize[0.61,0.74]} & 71.5 & +0.037$^{\ast}$ & 0.674 & +0.023 & 0.856 {\scriptsize[0.78,0.93]} & 87.1 & +0.055$^{\ast}$ \\
\rowcolor{rowhl}\quad + both (\textbf{full system}) & lead & 0.699 {\scriptsize[0.63,0.76]} & 72.4 & +0.058$^{\ast}$ & 0.696 & +0.045$^{\ast}$ & 0.867 {\scriptsize[0.78,0.94]} & 87.1 & +0.066$^{\ast}$ \\
\bottomrule
\end{tabular}

\end{table*}

\subsection{Overall}
\begin{figure}[t]
\centering
\includegraphics[width=\columnwidth]{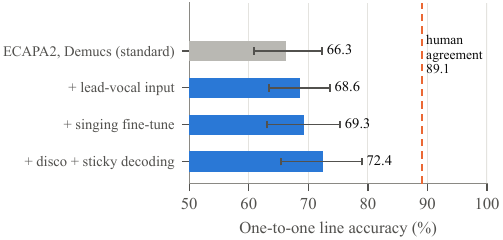}
\caption{ColorCoded-SD (all 1,740 lines): line accuracy as components are added cumulatively (error bars: 95\% song
bootstrap CI).}
\label{fig:progress}
\end{figure}

\Cref{tab:main} gives the main results and \cref{fig:progress} the cumulative effect of each component. On
ColorCoded-SD the full system improves the standard pipeline from 0.641 to 0.699 \bthree{} ($+0.058$, $p=0.001$) and
from 66.3\% to 72.4\% line accuracy ($p=0.007$); on confirmed lines alone the gain is $+0.045$ ($p=0.012$). On the
unseen IdolSongsJp singers it improves 0.800 to 0.867 ($+0.066$, $p=0.003$), and lead input with ECAPA2-ft and sticky
decoding reaches 0.875 (87.7\%). Against human agreement (89.1\%) the full system closes about a quarter of the gap
on ColorCoded-SD.

\subsection{Where the separator leaves a lead voice, the input matters most}
\label{sec:res-input}

\begin{figure}[t]
\centering
\includegraphics[width=\columnwidth]{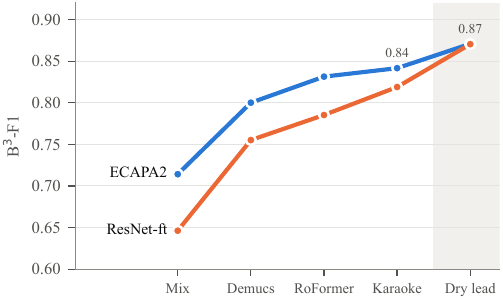}
\caption{IdolSongsJp (14 songs, all 309 lines): \bthree{} by input. Mix: no separation; Demucs: HT-Demucs vocals;
RoFormer: BS-RoFormer vocals; Karaoke: Mel-RoFormer lead vocals; Dry lead: each singer's own studio take (an oracle
separator). The gap between encoders closes as the input gets cleaner.}
\label{fig:input}
\end{figure}

\Cref{fig:input} varies only the front end on IdolSongsJp. Each step towards a cleaner lead helps both encoders, and
removing \emph{other voices} helps more than a better all-vocal separator: from Demucs, BS-RoFormer adds +0.03 and
the karaoke model +0.04 (ECAPA2) to +0.06 (ResNet-ft). The gap between the two encoders shrinks from 0.07 on the mix
to 0.02 with karaoke input and vanishes with the dry takes (0.871 for both), so with clean input the choice of encoder
stops mattering. On ColorCoded-SD, on the 1,308 lines that have both a lead and a Demucs voiceprint, lead input adds
+0.039 for ECAPA2 ($p=0.002$), +0.044 for ResNet-ft ($p=0.011$) and +0.059 for ECAPA2-ft ($p<0.001$). Scored on all
lines the effect is smaller (+0.020 for ECAPA2, $p=0.10$), because the lines the separator silences and the lines too
short for a voiceprint receive no benefit.

\subsection{Encoders}
\label{sec:res-enc}

\begin{table}[t]
\centering\small
\caption{Encoders on ColorCoded-SD (AHC, true $k$, all lines). EER: equal error rate of same/different-singer line
pairs within songs on Demucs vocals. Mix: no separation. CPU: one 3\,s window on one thread.}
\label{tab:enc}
\setlength{\tabcolsep}{2.6pt}
\begin{tabular}{@{}lrrcccc@{}}
\toprule
Encoder & Params & CPU & EER$\downarrow$ & Mix & Demucs & Lead \\
\midrule
ResNet (speech) & 15.4M & 0.13\,s & 30.7 & 0.557 & 0.573 & 0.595 \\
ECAPA2 (speech) & 27.1M & 1.1\,s & 27.8 & 0.551 & \textbf{0.641} & 0.661 \\
ResNet-ft & 15.4M & 0.13\,s & \textbf{26.5} & \textbf{0.595} & 0.634 & 0.665 \\
ECAPA2-ft & 27.1M & 1.1\,s & 26.7 & --- & 0.639 & \textbf{0.672} \\
\midrule
Both -ft, concat. & 42.5M & 1.2\,s & 25.7 & 0.591 & 0.664 & 0.696 \\
\bottomrule
\end{tabular}

\end{table}

Singing adaptation improves the ResNet over its speech initialization by +0.060 \bthree{} ($p<0.001$, Demucs input;
EER 30.7\%$\to$26.5\%) and makes it statistically equal to ECAPA2 ($-0.007$, $p=0.63$) at half the parameters and
1/8 of the CPU time (\cref{tab:enc}). Without separation, ResNet-ft is the most robust encoder: ECAPA2 loses 0.090
\bthree{} on the full mix and ResNet-ft 0.039, so ResNet-ft leads by +0.044 ($p=0.038$). Adapting ECAPA2 helps less
(+0.011 over ECAPA2 on lead input, $p=0.30$). On IdolSongsJp's unseen singers the adaptation transfers with clean
input: on the dry takes ResNet-ft beats its initialization by +0.038 ($p=0.049$). Concatenating the two adapted
encoders is the best per-song configuration (0.696 on lead input), at the cost of running both.

\subsection{Song-level decoding}
\label{sec:res-dec}

\begin{figure}[t]
\centering
\includegraphics[width=\columnwidth]{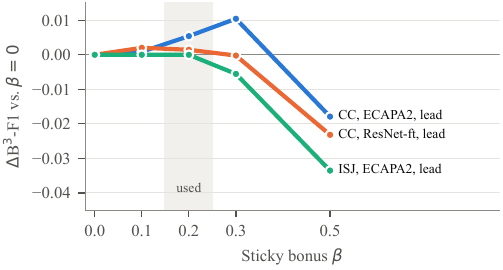}
\caption{Change in \bthree{} with the sticky bonus $\beta$ alone (lead input, true $k$; CC: ColorCoded-SD, ISJ:
IdolSongsJp). Alone, the sequence prior helps little; $\beta\ge0.5$ over-smooths.}
\label{fig:sticky}
\end{figure}

\begin{figure}[t]
\centering
\includegraphics[width=\columnwidth]{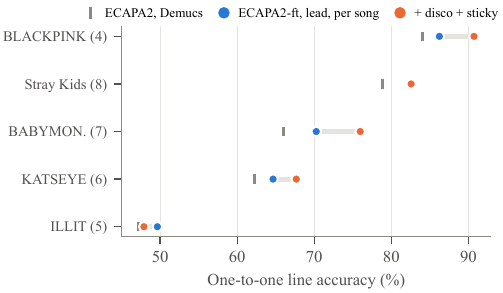}
\caption{Per-group line accuracy on ColorCoded-SD (all lines; members in parentheses; songs: BLACKPINK 11, KATSEYE
12, BABYMONSTER 11, ILLIT 6, Stray Kids 3). Linking helps groups with many songs; for Stray Kids it changes nothing
(markers coincide), and for ILLIT, whose members sound alike, it propagates errors.}
\label{fig:groups}
\end{figure}

Used alone, neither decoder is significant on ColorCoded-SD: sticky decoding adds +0.005 \bthree{} ($p=0.10$;
\cref{fig:sticky}) and discography linking +0.007 ($p=0.40$), although linking already improves line accuracy by
+2.2 points ($p=0.001$). Combined, with the sticky pass as the decoder inside linking, they add +0.028 \bthree{} and
+3.1 points of accuracy over ECAPA2-ft per song (both $p<0.001$), and +0.023 on confirmed lines ($p<0.001$). The gain
grows with the number of songs a group contributes (\cref{fig:groups}): +4.5 points for BLACKPINK, +5.7 for
BABYMONSTER and +3.0 for KATSEYE (11--12 songs each), none for Stray Kids (3 songs), and $-1.8$ for ILLIT, where
per-song clusters are least reliable. IdolSongsJp shows the limit of the assumption: its 18 session singers recombine
across songs rather than forming a fixed group, linking does not help ($-0.003$), and only the sticky pass adds a
small, non-significant gain (+0.016, $p=0.24$).

\subsection{Unknown singer count}
\label{sec:res-k}
Without a given $k$, a per-song distance threshold for AHC, chosen leave-one-group-out, finds the exact singer count
in only 23\% (ECAPA2, Demucs) and 28\% (ECAPA2-ft, lead) of ColorCoded-SD songs and over-counts by about one singer
(mean error 1.6 and 1.4); an eigengap criterion under-counts by 2.4. Because \bthree{} tolerates over-splitting it
hardly changes (0.674 vs.\ 0.672 for ECAPA2-ft on lead input), but one-to-one accuracy drops by 2.4--4.8 points.
Counting singers from audio remains open; in the application the member list provides $k$.

\subsection{Commercial speech diarization}
\label{sec:commercial}
We ran ElevenLabs Scribe v2~\citep{elevenlabs2025scribe}, a commercial speech-to-text service with speaker
diarization (threshold 0.1), on the full mixes and gave each ColorCoded-SD line its maximum-overlap speaker (or the
previous line's speaker when no word overlaps). Scribe finds 1.35 speakers per song on average, against 5.3 singers,
and a single speaker in 33 of 43 songs, so most lines collapse into one cluster: \bthree{} is 0.386 and one-to-one
accuracy 30.6\%. The standard pipeline of \cref{tab:main} improves on it by +0.255 \bthree{} and the full system by
+0.314 \bthree{} and +41.8 accuracy points (both $p<0.001$). Speech diarizers are tuned to separate voices that differ
much more than the members of one group, and Scribe also runs on the full mix rather than separated vocals.

\subsection{Error analysis}
\label{sec:res-err}

\begin{figure*}[t]
\centering
\includegraphics[width=\textwidth]{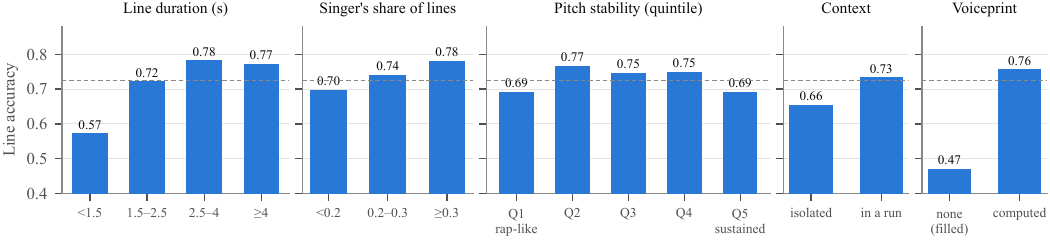}
\caption{Line accuracy of the full system by line attribute (ColorCoded-SD, all 1,740 lines; dashed: overall 0.72).
Pitch stability: fraction of voiced frames whose YIN $f_0$ stays within one semitone of the next frame. Voiceprint:
lines without any window (too short or silent) take the cluster of their nearest neighbour.}
\label{fig:errors}
\end{figure*}

\Cref{fig:errors} breaks down the errors of the full system (accuracy 0.72). The clearest factor is the amount of
audio: lines under 1.5\,s are 0.57 correct, and the 11\% of lines without a voiceprint, which inherit a neighbour's
cluster, are 0.47 correct against 0.76 for the rest. Minor members (under 20\% of a song's lines) and isolated lines
without a same-singer neighbour are also harder (0.70 and 0.66). Both ends of the pitch-stability range are hard
(0.69): rap-like lines and sustained hook lines, which are most often doubled by other members. Across groups,
accuracy ranges from 0.91 (BLACKPINK) to 0.48 (ILLIT), following how similar the members sound.

\subsection{Negative results}
\label{sec:negative}

\begin{table}[t]
\centering\small
\caption{Approaches tried during development that did not help. Numbers are from intermediate evaluations on the ColorCoded songs and are not comparable to \cref{tab:main}.}
\label{tab:neg}
\setlength{\tabcolsep}{3pt}
\begin{tabular}{@{}>{\raggedright\arraybackslash}p{0.47\columnwidth}>{\raggedright\arraybackslash}p{0.49\columnwidth}@{}}
\toprule
Approach & Outcome \\
\midrule
4$\times$ training data (577 $\to$ 2,277 songs) & no gain (LOO-centroid 79.8 vs.\ 80.5) \\
Lead-only training data (600 songs) & tie; $-$0.027 \bthree{} on IdolSongsJp \\
Closed-form weight edits (gain, ridge) & $\le$+0.014 \bthree{}, mostly n.s. \\
Within-line / group-level whitening & $-$0.01 to $-$0.02 \bthree{} \\
Song-context transformer & oracle-$k$ +1.8, LOO-centroid $-$2.5 \\
Distilling ECAPA2 into the ResNet & worse than using ECAPA2 \\
Training on the full mix & degraded after 250 steps \\
Fusing Demucs and lead voiceprints & $-$0.01 to $-$0.02 \bthree{} \\
Target-singer extraction loop & no gain in grouping \\
Multi-singer line detector & AUC 0.80; no all-line gain \\
\bottomrule
\end{tabular}
\end{table}

\Cref{tab:neg} lists approaches that did not help; two bound the headroom. First, encoder quality saturates early:
the same recipe on a quarter of the data reaches the same within-song accuracy, and training on lead-separated data
does not transfer better than training on Demucs vocals. Second, we tried to isolate each singer directly with a
target-singer extraction loop (cluster, enroll each cluster, extract, re-score). Extractors trained from scratch on
same-song mixtures used the enrollment only weakly; a pretrained vocal separator cannot be conditioned at its
bottleneck because its skip connections carry the mixture to the output (an oracle per-channel conditioning at every
encoder layer reaches only 3.0\,dB SNR); and fine-tuning a public singer-conditioned a cappella
separator\footnote{\url{https://huggingface.co/AcappellaSet/ACappellaSet-Models}} on our data learned to clean the lead
voice but ignored the singer cue (2.67 vs.\ 2.68\,dB SNR with the right and a wrong enrollment). Singer-selective
extraction for same-song voices remains open.

\section{Limitations}
\label{sec:limits}
\begin{itemize}
  \item \textbf{Tuning on the benchmark songs.} Components were developed and tuned on the ColorCoded songs, so
  ColorCoded-SD numbers for tuned components are mildly optimistic. IdolSongsJp is untouched and confirms the overall
  gain, but has only 14 songs and wide intervals.
  \item \textbf{Coverage.} Only single-singer lines are labelled (84\% of annotated lines; 45\% of IdolSongsJp vocal
  time), and 11\% of ColorCoded-SD lines are too short or quiet for a voiceprint. Line timings come from
  annotations, not forced alignment.
  \item \textbf{Domain.} ColorCoded-SD is K-pop and IdolSongsJp J-pop style; Western pop, solo artists with features
  and other languages are not evaluated.
  \item \textbf{Comparability.} Our within-song, line-level protocol differs from FruitsMusic's group-level,
  frame-level protocol~\citep{suda2024fruitsmusic}.
  \item \textbf{Cost.} The karaoke separator takes about 3.5\,min per song on a laptop GPU.
\end{itemize}

\section{Conclusion}
We introduced line-level singer diarization for multi-member pop songs, released the ColorCoded-SD benchmark with
alignment and scoring tools, and decomposed the error of a speaker-embedding pipeline. Isolating the lead voice is the
largest single factor wherever a lead voice survives separation, and with clean takes the encoder hardly matters;
song-level decoding exploits turn-taking and a group's recurring members; together they improve the standard pipeline
by +0.058 \bthree{} on ColorCoded-SD and +0.066 on unseen IdolSongsJp singers. The open problems are short lines,
lines the separator silences, multi-singer lines, and singer-selective extraction.

\section*{Ethics and Data Statement}
ColorCoded is a web application operated by nagiflow LLC, the author's company; its annotations are user-created
projects. The service's terms of service (revised effective December 24, 2026) permit their use for research and the
publication of de-identified datasets derived from them, with an opt-out for users. User identifiers were hashed and
no personal data is used. ColorCoded-SD is released at \url{https://huggingface.co/datasets/nagiflow/colorcoded-sd}
(public from December 24, 2026, when the revised terms take effect) under CC BY-NC 4.0. It contains official
recording identifiers (ISRC, MusicBrainz), 10\,Hz chroma fingerprints for alignment, line start and end times on a
0.25\,s grid, and pseudonymous singer labels consistent within a group; it contains no audio, lyric text, member
names, or user information. IdolSongsJp was used for non-commercial evaluation only, under its license, and its
derived labels are not redistributed; no model was trained on it. Singer identification could be misused to track
individuals; our task groups lines within commercially released songs and does not identify anyone outside them.

\bibliography{refs}

@inproceedings{thienpondt2023ecapa2,
  author    = {Thienpondt, Jenthe and Demuynck, Kris},
  title     = {{ECAPA2}: A Hybrid Neural Network Architecture and Training Strategy for Robust Speaker Embeddings},
  booktitle = {Proc. IEEE Automatic Speech Recognition and Understanding Workshop (ASRU)},
  year      = {2023}
}

@inproceedings{desplanques2020ecapa,
  author    = {Desplanques, Brecht and Thienpondt, Jenthe and Demuynck, Kris},
  title     = {{ECAPA-TDNN}: Emphasized Channel Attention, Propagation and Aggregation in {TDNN} Based Speaker Verification},
  booktitle = {Proc. Interspeech},
  year      = {2020}
}

@article{ravanelli2021speechbrain,
  author  = {Ravanelli, Mirco and others},
  title   = {{SpeechBrain}: A General-Purpose Speech Toolkit},
  journal = {arXiv preprint arXiv:2106.04624},
  year    = {2021}
}

@inproceedings{wang2023wespeaker,
  author    = {Wang, Hongji and Liang, Chengdong and Wang, Shuai and Chen, Zhengyang and Zhang, Binbin and Xiang, Xu and Deng, Yanlei and Qian, Yanmin},
  title     = {{WeSpeaker}: A Research and Production Oriented Speaker Embedding Learning Toolkit},
  booktitle = {Proc. IEEE ICASSP},
  year      = {2023}
}

@article{chen2022wavlm,
  author  = {Chen, Sanyuan and others},
  title   = {{WavLM}: Large-Scale Self-Supervised Pre-Training for Full Stack Speech Processing},
  journal = {IEEE Journal of Selected Topics in Signal Processing},
  volume  = {16},
  number  = {6},
  year    = {2022}
}

@inproceedings{li2024mert,
  author    = {Li, Yizhi and others},
  title     = {{MERT}: Acoustic Music Understanding Model with Large-Scale Self-supervised Training},
  booktitle = {Proc. ICLR},
  year      = {2024}
}

@inproceedings{rouard2023htdemucs,
  author    = {Rouard, Simon and Massa, Francisco and D{\'e}fossez, Alexandre},
  title     = {Hybrid Transformers for Music Source Separation},
  booktitle = {Proc. IEEE ICASSP},
  year      = {2023}
}

@inproceedings{lu2024bsroformer,
  author    = {Lu, Wei-Tsung and Wang, Ju-Chiang and Kong, Qiuqiang and Hung, Yun-Ning},
  title     = {Music Source Separation with Band-Split {RoPE} Transformer},
  booktitle = {Proc. IEEE ICASSP},
  year      = {2024}
}

@article{wang2023melroformer,
  author  = {Wang, Ju-Chiang and Lu, Wei-Tsung and Won, Minz},
  title   = {Mel-Band {RoFormer} for Music Source Separation},
  journal = {arXiv preprint arXiv:2310.01809},
  year    = {2023}
}

@misc{becruily2025karaoke,
  author       = {becruily},
  title        = {Mel-Band {RoFormer} Karaoke Model},
  howpublished = {\url{https://huggingface.co/becruily/mel-band-roformer-karaoke}},
  year         = {2025}
}

@inproceedings{khosla2020supcon,
  author    = {Khosla, Prannay and Teterwak, Piotr and Wang, Chen and Sarna, Aaron and Tian, Yonglong and Isola, Phillip and Maschinot, Aaron and Liu, Ce and Krishnan, Dilip},
  title     = {Supervised Contrastive Learning},
  booktitle = {Advances in Neural Information Processing Systems (NeurIPS)},
  year      = {2020}
}

@inproceedings{bagga1998bcubed,
  author    = {Bagga, Amit and Baldwin, Breck},
  title     = {Entity-Based Cross-Document Coreferencing Using the Vector Space Model},
  booktitle = {Proc. COLING-ACL},
  year      = {1998}
}

@article{amigo2009comparison,
  author  = {Amig{\'o}, Enrique and Gonzalo, Julio and Artiles, Javier and Verdejo, Felisa},
  title   = {A Comparison of Extrinsic Clustering Evaluation Metrics Based on Formal Constraints},
  journal = {Information Retrieval},
  volume  = {12},
  number  = {4},
  pages   = {461--486},
  year    = {2009}
}

@article{kuhn1955hungarian,
  author  = {Kuhn, Harold W.},
  title   = {The {Hungarian} Method for the Assignment Problem},
  journal = {Naval Research Logistics Quarterly},
  volume  = {2},
  number  = {1--2},
  pages   = {83--97},
  year    = {1955}
}

@article{landini2022vbx,
  author  = {Landini, Federico and Profant, J{\'a}n and Diez, Mireia and Burget, Luk{\'a}{\v{s}}},
  title   = {{Bayesian HMM} Clustering of x-vector Sequences ({VBx}) in Speaker Diarization: Theory, Implementation and Analysis on Standard Tasks},
  journal = {Computer Speech \& Language},
  volume  = {71},
  year    = {2022}
}

@article{park2022review,
  author  = {Park, Tae Jin and Kanda, Naoyuki and Dimitriadis, Dimitrios and Han, Kyu J. and Watanabe, Shinji and Narayanan, Shrikanth},
  title   = {A Review of Speaker Diarization: Recent Advances with Deep Learning},
  journal = {Computer Speech \& Language},
  volume  = {72},
  year    = {2022}
}

@inproceedings{plaquet2023powerset,
  author    = {Plaquet, Alexis and Bredin, Herv{\'e}},
  title     = {Powerset Multi-class Cross Entropy Loss for Neural Speaker Diarization},
  booktitle = {Proc. Interspeech},
  year      = {2023}
}

@inproceedings{medennikov2020tsvad,
  author    = {Medennikov, Ivan and others},
  title     = {Target-Speaker Voice Activity Detection: A Novel Approach for Multi-Speaker Diarization in a Dinner Party Scenario},
  booktitle = {Proc. Interspeech},
  year      = {2020}
}

@inproceedings{suda2024fruitsmusic,
  author    = {Suda, Hitoshi and Yoshida, Shunsuke and Nakamura, Tomohiko and Fukayama, Satoru and Ogata, Jun},
  title     = {{FruitsMusic}: A Real-World Corpus of {Japanese} Idol-Group Songs},
  booktitle = {Proc. International Society for Music Information Retrieval Conference (ISMIR)},
  year      = {2024},
  note      = {arXiv:2409.12549}
}

@inproceedings{suda2025idolsongsjp,
  author    = {Suda, Hitoshi and others},
  title     = {{IdolSongsJp} Corpus: A Multi-Singer Song Corpus in the Style of {Japanese} Idol Groups},
  booktitle = {Proc. International Society for Music Information Retrieval Conference (ISMIR)},
  pages     = {647--654},
  year      = {2025},
  note      = {arXiv:2507.01349}
}

@article{xu2026singerinformed,
  author  = {Xu, Jocelyn and Kim, Minje},
  title   = {Singer-Informed Vocal Source Separation for Multi-Singer Music Mixtures},
  journal = {arXiv preprint arXiv:2608.14516},
  year    = {2026}
}

@misc{elevenlabs2025scribe,
  author       = {{ElevenLabs}},
  title        = {Scribe v2 speech-to-text with speaker diarization},
  howpublished = {\url{https://elevenlabs.io/speech-to-text}},
  note         = {Accessed November 2026},
  year         = {2025}
}

\end{document}